\documentclass[12pt]{article}
\usepackage{amsfonts,amssymb}
\usepackage{color}
\usepackage{epic}
\usepackage{graphics}

\def\lddots{\mathinner{\mkern1mu\raise1pt\hbox{.}\mkern2mu
\raise4pt\hbox{.}\mkern2mu\raise7pt\vbox{\kern7pt\hbox{.}}\mkern1mu}}
\makeatletter
\def\numberbysection{\@addtoreset{equation}{section}
\def\theequation{\thesection.\arabic{equation}}}
\makeatother

\numberbysection

\newcommand{\be}{\begin{eqnarray}}
\newcommand{\ee}{\end{eqnarray}}
\newcommand{\non}{\nonumber}

\begin{document}

\begin{titlepage}
\vskip 0.4cm
\strut\hfill
\vskip 0.8cm
\begin{center}


{\bf {\Large Integrable boundary conditions and modified Lax equations}}

\vspace{10mm}

{\large Jean Avan\footnote{avan@ptm.u-cergy.fr}$^{a}$ and Anastasia Doikou\footnote{doikou@bo.infn.it}$^{b}$}

\vspace{10mm}

{\small $^a$ LPTM, Universite de Cergy-Pontoise (CNRS UMR 8089), Saint-Martin 2\\
2 avenue Adolphe Chauvin, F-95302 Cergy-Pontoise Cedex, France}

{\small $^b$ University of Bologna, Physics Department, INFN Section \\
Via Irnerio 46, Bologna 40126, Italy}

\end{center}

\vfill

\begin{abstract}

We consider integrable boundary conditions for both discrete and
continuum classical integrable models. Local
integrals of motion generated by the corresponding ``transfer''
matrices give rise to time evolution equations for the
initial Lax operator. We systematically identify the modified Lax
pairs for both discrete and continuum boundary integrable models,
depending on the classical $r$-matrix and the boundary matrix.

\end{abstract}

\vfill
\baselineskip=16pt

\end{titlepage}

\section{Introduction}

Lax representation of classical dynamical evolution equations
\cite{lax} is one key ingredient in the modern theory of classical
integrable systems \cite{GGKM}--\cite{BBT} together with
the associated notion of classical $r$-matrix \cite{skl, sts}.
It takes the generic form of an isospectral evolution equation:
${d L\over d t} = [L,\ A]$,
where $L$ encapsulates the dynamical variables and $A$ defines the time evolution.
We shall generically consider the situation where $L$ and $A$ depend on a complex (spectral)
parameter.
The spectrum of the Lax matrix or its extension
(transfer matrices), or equivalently the invariant coefficients of
the characteristic determinant, thus provide automatically candidates
to realize the hierarchy of Poisson-commuting Hamiltonians
required by Liouville's theorem \cite{liouv, arn}. Existence of the
classical $r$-matrix then guarantees Poisson-commutativity of these
natural dynamical quantities taken as generators of the algebra of
classical conserved charges.

The question of finding an appropriate time evolution matrix $A$
for a given Lax matrix $L$ therefore entails the possibility of systematically
constructing classically integrable models once a Lax matrix
is defined with suitable properties. Depending on which properties
are emphasized, this problem may be approached in several ways.
One approach uses postulated Poisson algebra properties of $L$, specifically
the $r$-matrix structure, as a starting point, and establishes a systematic
construction of the time evolution
operator $A$ associated to the Hamiltonian evolution obtained
from any function in the enveloping algebra of
the Poisson-commuting traces of such an $L$ matrix. Such a formulation
was proposed long time ago
\cite{skl, sts} for the bulk case, when the Poisson structure for
$L$ is a simple linear or quadratic $r$-matrix structure.

We shall consider here more general
situations when the dynamical system also depends on supplementary
parameters encapsulated into a matrix $K$. We shall restrict ourselves
to the situation where these parameters are non-dynamical, i.e.
the Poisson brackets with themselves and with the initial
``bulk'' parameters is zero.  For situations where
the extra parameters are dynamical see \cite{BasDel,paper}. Note that interpretation of the physical
meaning of these $c$-number parameters will come a posteriori when computing the associated
Hamiltonians. In particular any physical interpretation of the
$K$ matrix as a description of the ``boundary properties''
(external fields, ...) may not be appropriate
in all cases as shall appear in our discussion of examples.
We shall however keep this designation as a book-keeping device throughout this paper.

Our central purpose will be twofold. We shall first of all define generic sets of sufficient algebraic
conditions on $K$ also formulated in terms of the ``bulk'' $r$-matrix. A
``boundary--modified'' generating matrix of
candidate conserved quantities, hereafter denoted ${\cal T}$,
will be accordingly constructed as a suitable combination of $L$ and $K$ matrices.
The idea for such a construction naturally arises when considering
the semi-classical limit of the well--known Cherednik-Sklyanin
reflection algebras preserving bulk quantum integrability
\cite{sklyanin}. We shall then redefine
the time evolution operator $A$ associated to a given
Hamiltonian constructed from ${\cal T}$ (modified monodromy
matrix) from the new basic elements,
i.e. the matrix ${\cal T}$, the bulk $r$-matrix or $(r,s)$
pair \cite{sts, maillet}, and the reflection
matrix $K$. Such a construction was exemplified in \cite{durham}; what
we propose here is however a generic, systematic procedure to obtain modified
``boundary'' or ``folded'' classical integrable systems from initial
pure ``bulk'' systems. Note also that the example worked out in \cite{durham}
is precisely a case which the $K$-matrix does describe boundary effects.

It must be emphasized here that an alternative, analytic approach
to this question, at least in the bulk case,
was extensively described in \cite{BBT} (chapter 3). In this approach one
uses instead the analytic
properties (location of poles and algebraic structure of residues)
of the Lax matrix $L$ as a meromorphic function of the spectral parameter.
They provide for a unique consistent form of an associated
$A$ matrix. The subsequent Lax equation is then developed into separate
equations corresponding to the poles of $L$ and $A$. The Poisson structure and $r$
matrix structure are only then defined as providing a consistent
Hamiltonian interpretation of this Lax equation.
Similar reformulation of our results must exist but we shall
not consider them here.

We shall expand here the $r$-matrix approach in three situations. We
start with the simpler case where the initial bulk structure is a
linear Poisson structure for a Lax matrix parametrized by a
single $r$-matrix. We then develop the cases of both discrete and
continuous parametrized Lax matrices relevant to the description of
systems on a lattice or on a continuous line. In these last two
situations the relevant $r$-matrix structure is a quadratic Sklyanin-type
bracket \cite{kulishsklyanin}.
We shall in all three
cases derive (or actually rederive, in some cases) the
form of the generating functional for
Poisson-commuting Hamiltonians, and establish the explicit general
formula yielding the $A$ operator. Explicit examples shall be developed
in all cases, albeit restricting
ourselves to the simplest situations of non--dynamical non--constant
double--pole rational $r$-matrices. More complicated situations
(trigonometric and/or dynamical $r$-matrices, relevant
for e.g. sine-Gordon models or affine Toda field theories) will be left for further
studies.

\section{Linear Poisson structure}

We consider here the original situation \cite{lax} where the full dynamical
system under study is represented by a single Lax matrix, living
in a representation of a finite-dimensional Lie algebra or a loop
algebra. In this last case the Lax matrix also depends on a complex parameter
$\lambda$ known as ``spectral parameter''.
In all cases the requirement of Poisson-commutativity for candidate
Hamiltonians $Tr L^n$ is equivalent \cite{babelon} to the
existence of a classical $r$-matrix \cite{skl, sts} realizing a linear
Poisson structure for $L$. Indeed, consider the Lax pair $(L,\ A)$
satisfying \be {\partial L \over
\partial t}  = \Big [A,\  L \Big ] \ee the associated spectral problem
\be L(\lambda)\ \psi = u\ \psi, ~~~~~~ \det (L(\lambda) -u) =0
\label{spectral2} \ee provides the integrals of
motion, obtained through the expansion in powers
of the spectral parameter $\lambda$ of $tr L(\lambda)$.
Alternatively, in particular when no spectral parameter exists, one
should consider the traces of powers of the Lax matrix $tr L^n (\lambda)$
as natural Hamiltonians.

Assuming that the $L$
matrix satisfies the fundamental relation \be \Big
\{L_a(\lambda),\ \ L_b(\mu) \Big \} = \Big [ r_{ab}(\lambda-\mu),\
L_a(\lambda) +L_b(\mu)  \Big ] \label{fundam} \ee it is shown using (\ref{fundam}) that
for any integer $n,m$: \be
\Big \{ trL^n(\lambda),\ tr L^m(\mu) \Big \} =0. \ee
The reciprocal property was shown in \cite{babelon}.
Let us recall \cite{skl, sts, babelon}
how one may identify the $A$-operator associated to the various
charges in involution. Bearing in mind the fundamental relation
(\ref{fundam}) it is shown that \be \Big \{ tr_a\
L_a^n(\lambda),\ L_b(\mu) \Big \} = n\ tr_a\ \Big
(L_a^{n-1}(\lambda)\ r_{ab}(\lambda -\mu)\Big )\ L_b(\mu) - n\
L_b(\mu)\ tr_a\ \Big (L_a^{n-1}(\lambda) r_{ab}(\lambda -\mu)\Big
). \label{lax2} \ee From (\ref{lax2}) one extracts $A$:
\be A_n(\lambda,\ \mu) = n\ tr_a\ \Big (L_a^{n-1}(\lambda)
r_{ab}(\lambda -\mu)\Big ). \ee
In the case of the simplest rational non-dynamical
$r$-matrices \cite{young}
\be r(\lambda) = {{\mathbb P} \over \lambda}
~~~~\mbox{where} ~~~~{\mathbb P}=\sum_{i,j=1}^{N} E_{ij} \otimes
E_{ji} \label{rr} \ee ${\mathbb P}$ is the permutation operator,
and $~(E_{ij})_{kl} = \delta_{ik} \delta_{jl}$, we end up with a simple form for $A_n$:
\be A_n(\lambda,\ \mu) =  {n
\over \lambda -\mu} L^{n-1}(\lambda) \ee and as usual to obtain
the Lax pair associated to each local integral of motion one has
to expand $A_n = \sum_i {A_n^{(i)} \over \lambda^i}$. Note that generically,
using the dual formulation of the classical $r$-matrix \cite{sts}
one also has:
\be A(\lambda, \mu) = Tr (r(\lambda, \mu) dH) \ee
where $H$ is the Hamiltonian expressed as any function in the
enveloping algebra generated by $Tr L^n$.

Ultimately we would like to consider an extended classical algebra
in analogy to the quantum boundary algebras arising in integrable
systems with non-trivial boundary conditions that preserve
integrability. Subsequently we shall deal with two types of
algebras which may be associated with the two types of known
quantum boundary conditions. These boundary conditions are known
as soliton preserving (SP), traditionally studied in the framework
of integrable quantum spin chains (see e.g. \cite{sklyanin},
\cite{dvg}--\cite{masa}), and soliton non-preserving (SNP)
originally introduced in the context of classical integrable field
theories \cite{durham}, and further investigated in \cite{gand,
dema}. SNP boundary conditions have been also introduced and
studied for integrable quantum lattice systems
\cite{doikousnp}--\cite{crdo}. From the algebraic perspective the
two types of boundary conditions are associated with two distinct
algebras, i.e. the reflection algebra \cite{sklyanin} and the
twisted Yangian respectively \cite{molev, moras} (see also 
\cite{dema, doikouy, crdo, mac, doikounp}). The classical
versions of both algebras will be defined subsequently in the text
(see section 3.2). It will be convenient for our purposes here to
introduce some useful notation: \be && \hat r_{ab}(\lambda) =
r_{ba}(\lambda) ~~~\mbox{for SP}, ~~~~\hat r_{ab}(\lambda) =
r_{ba}^{t_a t_b}(\lambda) ~~~\mbox{for SNP} \non\\ &&
r^*_{ab}(\lambda) =r_{ab}(\lambda) ~~~\mbox{for SP}, ~~~~
r^*_{ab}(\lambda) = r_{ba}^{t_b}(-\lambda) ~~~\mbox{for SNP}
\non\\ && \hat r^*_{ab}(\lambda) =r_{ba}(\lambda) ~~~\mbox{for
SP}, ~~~~\hat r^*_{ab}(\lambda) = r_{ab}^{t_a}(-\lambda)
~~~\mbox{for SNP} \label{notation00} \ee together with: \be \hat
L(\lambda) =-L(-\lambda) ~~~\mbox{for SP}, ~~~~\hat L(\lambda) =
L^t(-\lambda) ~~~\mbox{for SNP}. \label{auto} \ee In addition the
``boundary conditions'', to be interpreted on specific examples,
are parametrized by a single non-dynamical matrix ${\mathrm
k}(\lambda)$. We propose here the following set of algebraic
relations: \be  && \Big \{ {\cal T}_1(\lambda_1),\ {\cal
T}_2(\lambda_2) \Big \} = {\mathrm r}_{12}^-(\lambda_1,
\lambda_2)\ {\cal T}_1(\lambda_1) - {\cal T}_1(\lambda_1)\ \tilde
{\mathrm r}_{12}^-(\lambda_1, \lambda_2) \non\\ && + {\mathrm
r}_{12}^+(\lambda _1, \lambda_2)\ {\cal T}_2(\lambda_2) - {\cal
T}_2(\lambda_2)\ \tilde {\mathrm r}^{+}_{12}(\lambda_1, \lambda_2)
\label{13} \ee where we define: \be {\mathrm r}_{12}^-(\lambda_1,
\lambda_2) = r_{12}(\lambda_1 -\lambda_2)\ {\mathrm
k}_2(\lambda_2) -{\mathrm k}_2(\lambda_2)\
r^*_{12}(\lambda_1 +\lambda_2)  \non\\
\tilde {\mathrm r}_{12}^-(\lambda_1, \lambda_2) = {\mathrm
k}_2(\lambda_2)\ \hat r_{12}(\lambda_1 -\lambda_2) - \hat
r^*_{12}(\lambda_1 +\lambda_2)\
{\mathrm k}_2(\lambda_2) \non\\
{\mathrm r}_{12}^+(\lambda_1, \lambda_2) = r_{12}(\lambda_1
-\lambda_2)\ {\mathrm k}_1(\lambda_1) +{\mathrm k}_1(\lambda_1)\
\hat r^*_{12}(\lambda_1 +\lambda_2) \non\\ \tilde {\mathrm
r}_{12}^+(\lambda_1, \lambda_2) = {\mathrm k}_1(\lambda_1)\ \hat
r_{12}(\lambda_1 -\lambda_2)  + r^*_{12}(\lambda_1 +\lambda_2)\
{\mathrm k}_1(\lambda_1), \ee for a  ${\mathrm k}$ matrix satisfying:
\be && \Big \{ {\mathrm k}_1(\lambda),\ {\mathrm
k}_2(\mu) \Big \}=0, ~~~~~\Big \{{\mathrm k}_1(\lambda),\ L_2(\mu) \Big \} =0, \label{11}
\\ && r_{12}(\lambda_1 -\lambda_2)\ {\mathrm k}_1(\lambda)\
{\mathrm k}_2(\lambda_2) + {\mathrm k}_1(\lambda_1)\ \hat
r^{*}_{12}(\lambda_1+\lambda_2) {\mathrm k}_2(\lambda_2) \non\\ &&
= {\mathrm k}_1(\lambda_1)\ {\mathrm k}_2(\lambda_2)\ \hat
r_{12}(\lambda_1 -\lambda_2) + {\mathrm k}_2(\lambda_2)\
r^*_{12}(\lambda_1+\lambda_2)\ {\mathrm k}_1(\lambda_1). \label{12} \ee
We then show:
\\
\\
{\it \underline{Theorem 2.1.}}: The quantity \be && {\cal T}(\lambda) = L(\lambda)\ {\mathrm
k}(\lambda) + {\mathrm k}(\lambda)\ \hat L(\lambda) \label{repr}
\ee is a representation of the algebra defined by
(\ref{13}), with ${\mathrm k}$ obeying (\ref{11}),
(\ref{12}).
\\
\\
{\it \underline{Proof}}: We shall need in addition
to (\ref{fundam}) the following exchange relations: \be && \Big
\{\hat L_a(\lambda),\  L_b(\mu)  \Big \} = \hat L_a(\lambda) \hat
r_{ab}^*(\lambda +\mu) + \hat r^*_{ab}(\lambda+\mu) L_b(\mu) -\hat
r^*_{ab}\lambda+\mu) \hat L_a(\lambda) - L_b(\mu) \hat
r^*_{ab}(\lambda+\mu)
\non\\
&& \Big \{ L_a(\lambda),\ \hat  L_b(\mu)  \Big \} = L_a(\lambda)
r_{ab}^*(\lambda +\mu) + r_{ab}^*(\lambda +\mu) \hat L_b(\mu) -
r_{ab}^*(\lambda +\mu)
L_a(\lambda)- \hat L_b(\mu) r_{ab}^*(\lambda +\mu) \non\\
&&  \Big \{\hat L_a(\lambda),\ \hat  L_b(\mu)  \Big \} =  \Big
[\hat r_{ab}(\lambda-\mu),\  \hat L_a(\lambda)  + \hat L_b(\mu)
\Big ]. \label {fundam2} \ee By explicit use of the
algebraic relations (\ref{12}) and (\ref{fundam2}) we obtain: \be
&& \Big \{ L_a(\lambda){\mathrm k}_a(\lambda)  + \hat L_a(\lambda)
{\mathrm k}_a(\lambda) ,\ L_b(\mu) {\mathrm
k}_b(\mu)+ {\mathrm k}_b(\mu)\hat L_b(\mu) \Big \} = \ldots \non\\
&& = {\mathrm r}_{ab}^-(\lambda, \mu) \Big ( L_a(\lambda){\mathrm
k}_a(\lambda) + {\mathrm k}_a(\lambda) \hat L_a(\lambda) \Big ) -
\Big ( L_a(\lambda){\mathrm k}_a(\lambda) +{\mathrm k}_a(\lambda)
\hat L_a(\lambda) \Big )\tilde {\mathrm r}_{ab}^-(\lambda, \mu)
\non\\   && + {\mathrm r}_{ab}^+(\lambda, \mu) \Big (
L_b(\mu){\mathrm k}_b(\mu) +{\mathrm k}_b(\mu) \hat L_b(\mu) \Big
) - \Big (L_b(\mu){\mathrm k}_b(\mu) +{\mathrm k}_b(\mu) \hat
L_b(\mu) \Big )\tilde {\mathrm r}_{ab}^+(\lambda, \mu) \non\\ && +
\Big ( L_a(\lambda) +L_b(\mu) \Big ) \Big
(-r_{ab}(\lambda-\mu){\mathrm k}_a(\lambda){\mathrm k}_b(\mu)
-{\mathrm k}_a(\lambda)\hat r^*_{ab}(\lambda+ \mu){\mathrm
k}_b(\mu) \non\\ && +{\mathrm k}_b(\mu)r_{12}^*(\lambda +\mu)
{\mathrm k}_a(\lambda) + {\mathrm k}_a(\lambda) {\mathrm k}_b(\mu)
\hat r_{12}(\lambda -\mu) \Big ) \non\\ && + \Big
(-r_{ab}(\lambda-\mu){\mathrm k}_a(\lambda){\mathrm k}_b(\mu)
-{\mathrm k}_a(\lambda)\hat r^*_{ab}(\lambda+ \mu){\mathrm
k}_b(\mu) \non\\ && +{\mathrm k}_b(\mu)r_{12}^*(\lambda +\mu)
{\mathrm k}_a(\lambda) + {\mathrm k}_a(\lambda) {\mathrm k}_b(\mu)
\hat r_{12}(\lambda -\mu) \Big ) \Big (\hat  L_a(\lambda) + \hat
L_b(\mu) \Big ). \label{fin} \ee Bearing however in mind that the
${\mathrm k}$-matrix obeys (\ref{12}) we conclude that the last
four lines of the equation above disappear, which shows
that (\ref{repr}) satisfies (\ref{13}), and this concludes our
proof. $\blacksquare$

Define ${\mathbb T}(\lambda) = {\mathrm k}^{-1}(\lambda)\ {\cal
T}(\lambda)$, we now prove:
\\
\\ {\it \underline{Theorem 2.2.}}:
\be \Big \{ tr_a {\mathbb
T}_a^N(\lambda),\ tr_b {\mathbb T}^M_b(\mu) \Big \} = 0. \label{th2} \ee
\\
{\it \underline{Proof}}: \be \Big \{ tr_a {\mathbb
T}_a^N(\lambda),\ tr_b {\mathbb T}^M_b(\mu) \Big \} = \sum_{n, m}
tr_{ab}\ {\mathbb T}_a^{N-n}(\lambda) {\mathbb T}_b^{M-m}(\mu)
\Big \{ {\mathbb T}_a(\lambda),\ {\mathbb T}_b(\mu) \Big \}
{\mathbb T}_a^{n-1}(\lambda) {\mathbb T}_b^{m-1}(\mu) \ee
employing (\ref{13}) the preceding expression becomes \be && \dots
\propto \non\\ && tr_{ab}  {\mathbb T}^{N-1}_a(\lambda) {\mathbb
T}^{M-1}_{b}(\mu) {\mathrm k}_a^{-1} {\mathrm k}_b^{-1} \Big (
{\mathrm r}_{ab}^{-} (\lambda, \mu) {\cal T}_a(\lambda) - {\cal
T}_a(\lambda) \tilde {\mathrm r}_{ab}^{-} (\lambda, \mu) +
{\mathrm r}_{ab}^{+} (\lambda, \mu){\cal T}_b(\mu)-{\cal T}_b(\mu)
\tilde {\mathrm r}_{ab}^{+} (\lambda, \mu) \Big ) \non\\ &&=
\ldots =0. \ee Note that in order to show that the latter
expression is zero we moved appropriately the factors in the
products within the trace and we used (\ref{12}). $\blacksquare$

We finally identify the modified Lax formulation associated
to the generalized algebra (\ref{11})--(\ref{13}) as:
\\
\\
{\it \underline{Theorem 2.3.}}:
Defining Hamiltonians as: $tr_a {\mathbb T}^n(\lambda) =
\sum_i{{\cal H}_n^{(i)} \over \lambda^i}$ the
classical equations of motion for ${\mathbb T}$: \be {\dot {\mathbb
T}}(\mu)= \Big \{{\cal H}_n^{(i)},\ {\mathbb T}(\mu) \Big \}
\label{cleq} \ee take a zero
curvature form  \be
{\dot {\mathbb T}}(\mu) =  \Big [{\mathbb A}(\lambda, \mu),\
{\mathbb T}(\mu) \Big], \ee where ${\mathbb A}_n$ is
identified as: \be {\mathbb A}_n(\lambda, \mu)  =n\  tr_a \Big (
{\mathbb T}_a^{n-1}(\lambda) {\mathrm k}_a^{-1}(\lambda) \tilde
{\mathrm r}^+_{ab}(\lambda, \mu) \Big ). \label{final3} \ee
\\
{\it \underline{Proof}}: \be && \Big
\{tr_a {\mathbb T}^n_a(\lambda),\ {\mathbb T}_b(\mu) \Big \}
=\ldots= \non\\ && n\ tr_a \Big ( {\mathbb T}_a^{n-1}(\lambda)
{\mathrm k}_a^{-1}(\lambda) {\mathrm k}_b^{-1}(\mu) ({\mathrm
r}^-_{ab}(\lambda, \mu) {\cal T}_a(\lambda) -{\cal
T}_a(\lambda)\tilde {\mathrm r}^-_{ab}(\lambda,\mu) +{\mathrm
r}^+_{ab}(\lambda, \mu) {\cal T}_b(\mu) -{\cal T}_b(\mu) \tilde
{\mathrm r}^+_{ab}(\lambda,\mu)) \Big ) \non\\ &&= n\ tr_a \Big (
{\mathbb T}_a^{n-1} {\mathrm k}_a^{-1}(\lambda) {\mathrm
k}_b(\mu)({\mathrm r}^{-}_{ab}(\lambda, \mu){\cal T}_a(\lambda)
-{\cal T}_a(\lambda) \tilde {\mathrm r}^{-}_{ab}(\lambda, \mu) )
\Big ) \non\\ && + n\ tr_a\Big ({\mathbb T}_a^{n-1} {\mathrm
k}_a^{-1}(\lambda) {\mathrm k}_b(\mu)({\mathrm
r}^{+}_{ab}(\lambda, \mu){\cal T}_b(\mu) -{\cal T}_b(\mu) \tilde
{\mathrm r}^{+}_{ab}(\lambda, \mu))  \Big ). \ee Taking into
account (\ref{12}) we see that the first term of RHS of the
equality above disappears and the last term may be appropriately
rewritten such as: \be \Big \{tr_a {\mathbb T}^n_a(\lambda),\
{\mathbb T}_b(\mu) \Big \} = n\ tr_a \Big ({\mathbb
T}^{n-1}(\lambda){\mathrm k}_a^{-1}(\lambda) \tilde {\mathrm
r}^{+}_{ab}(\lambda, \mu) \Big )\ {\mathbb T}_{b}(\mu) -n\
{\mathbb T}_{b}(\mu)\  tr_a \Big ( {\mathbb T}^{n-1}(\lambda)
{\mathrm k}_a^{-1}(\lambda) \tilde {\mathrm r}^{+}_{ab}(\lambda,
\mu) \Big ). \non\\ \ee From the latter formula (\ref{final3}) is
deduced. $\blacksquare$

Finally, let ${\cal T}(\lambda),\ {\cal T}'(\lambda)$ be two
representations of (\ref{13}), and let also \be \Big \{ {\cal
T}_a(\lambda),\ {\cal T}'_b(\mu) \Big \} =0. \label{no} \ee
It is then straightforward to show, based solely on the fact that
${\cal T},\ {\cal T}'$ satisfy (\ref{13}) and (\ref{no}) , that
the sum ${\cal T}(\lambda) +{\cal T}'(\lambda)$ is also a
representation of (\ref{13}).

\subsection{Examples}

We shall present here a simple example, starting from the classical
rational Gaudin model \cite{gaudin}. Details on the so called
``dual'' description of the Toda chain \cite{moeb, toda} and the DST
model \cite{sklyaninblac, kundubas}, both associated to the
$A_{{\cal N}-1}^{(1)}$ $r$-matrix \cite{jimbo}, will be presented
in a forthcoming publication. Before we proceed with the
particular example let us rewrite ${\mathbb A}_n$. We consider
a situation where the Poisson brackets
are parametrized by the simplest rational non-dynamical $r$-matrix
\cite{young}. After
substituting the rational $r$-matrix in (\ref{final3})
we get for both types of boundary conditions: \be && {\mathbb
A}_n(\lambda, \mu) = n\ \Big ({ {\mathbb T}^{n-1}(\lambda) \over
\lambda -\mu} + {{\mathrm k}(\lambda){\mathbb
T}^{n-1}(\lambda){\mathrm k}^{-1}(\lambda) \over \lambda +\mu}
\Big ) ~~~\mbox{for SP}, \non\\ && {\mathbb A}_n(\lambda, \mu) =
n\ \Big ({{\mathbb T}^{n-1}(\lambda) \over \lambda -\mu} -
{({\mathrm k}(\lambda){\mathbb T}^{n-1}(\lambda){\mathrm
k}^{-1}(\lambda))^t\over \lambda +\mu} \big ) ~~~\mbox{for
SNP}.\ee The $L$-matrix associated to the classical $gl_{{\cal
N}}$ Gaudin model, and satisfying the algebraic relation
(\ref{fundam}) is \be L(\lambda) =
\sum_{\alpha, \beta=1}^{{\cal N}} \sum_{i=1}^N {{\cal P}_{\alpha
\beta}^{(i)} \over \lambda -z_i} E_{\alpha \beta}  \ee where
${\cal P}_{\alpha \beta} \in gl_{{\cal N}}$. Recall that in the
``bulk'' case the integrals of motion are obtained through $tr\
L^n(\lambda)$. The first non-trivial
example reads \be t^{(2)} (\lambda) = tr\ L^2(\lambda) =
\sum_{\alpha, \beta=1}^{{\cal N}} \sum_{i, j=1}^{N} {{\cal
P}^{(i)}_{\alpha \beta} {\cal P}_{\beta \alpha}^{(j)} \over
(\lambda -z_i)(\lambda -z_j)} \ee and by taking the residue of the
latter expression for $\lambda \to z_i$ we obtain the Gaudin
Hamiltonian: \be H^{(2)} = \sum_{i \neq j=1}^N \sum_{\alpha, \beta
=1}^{{\cal N}} {{\cal P}_{\alpha \beta}^{(i)} {\cal P}_{\beta
\alpha}^{(j)} \over z_i -z_j}. \ee Let us now come to the generic
algebra (\ref{13}), (\ref{11}), (\ref{12}) considering both SP and SNP cases.
Based on the definitions (\ref{auto}) we have: \be && \hat
L(\lambda) = \sum_{\alpha, \beta =1}^{{\cal N}} \sum_{i=1}^N
{{\cal Q}_{\alpha \beta}^{(i)} \over \lambda +z_i} E_{\alpha}
~~~\mbox{where} \non\\ && {\cal Q}_{\alpha \beta}= {\cal
P}_{\alpha \beta} ~~~\mbox{for SP}, ~~~~{\cal Q}_{\alpha \beta}= -
{\cal P}_{\beta \alpha} ~~~\mbox{for SNP}. \ee Let us take as
a representation ${\cal T}(\lambda)= L(\lambda){\mathrm
k}(\lambda) +{\mathrm k}(\lambda) \hat L(\lambda) +K(\lambda)$
where ${\mathrm k}$ is a solution of (\ref{11}), (\ref{12}), and $K$
is a $c$-number representation of (\ref{13}) with zero Poisson bracket.
To obtain the relevant Hamiltonian
we now formulate $tr_a{\mathbb
T}^2(\lambda)$ (where ${\mathbb T} = k^{-1}{\cal T}$ and we also
define $\tilde K = {\mathrm k}^{-1}K$), and the Hamiltonian arises
as the residue of the latter expression at $\lambda = z_i$: \be
{\cal H}^{(2)} = \sum_{i\neq j=1}^N \sum_{\alpha, \beta=1}^{{\cal
N}} {{\cal P}_{\alpha \beta}^{(i)} {\cal P}_{\beta \alpha}^{(j)}
\over z_i -z_j} + \sum_{i, j=1}^N \sum_{\alpha, \beta, \gamma,
\delta, \epsilon=1}^{{\cal N}} {{\mathrm k}_{\alpha \gamma}
{\mathrm k}_{\delta \epsilon}{\cal P}_{\gamma \delta}^{(i)} {\cal
Q}_{\epsilon \alpha}^{(j)} \over z_i +z_j} + \sum_{i=1}^N
\sum_{\alpha, \beta, \gamma, \delta, \epsilon =1}^{{\cal N}}
\tilde K_{\epsilon \alpha} {\mathrm k}^{-1}_{\alpha \gamma} {\mathrm
k}_{\delta \epsilon} {\cal P}^{(i)}_{\gamma \delta} \label{h2} \ee
A special case of the generic algebra
(\ref{13}) is discussed in \cite{Hikami, crampe}.
Note also that expression (\ref{h2}) may be also
obtained as the semiclassical limit of the quantum $gl_{{\cal N}}$
inhomogeneous open spin chain for special boundary conditions,
(see e.g. \cite{gaudinb}), $z_j$ being the
inhomogeneities attached to each site $j$. However, here the
Hamiltonians are obtained directly from the
representations of our new classical algebra (\ref{13}), (\ref{11}),
(\ref{12}). It appears in this example that the parameters ${\mathrm k}(\lambda)$
play the role of ``coupling constants'' consistent with the integrable
``folding'' of a $2N$ site Gaudin model, and not the role of boundary parameters.

\section{Quadratic Poisson structures: the discrete case}

Quadratic Poisson structures first appeared as the well-known
Sklyanin bracket \cite{sklb}. A more general form, characterized
by a pair of respectively skew symmetric and symmetric matrices $(r,\ s)$
appeared in \cite{maillet} in the formulation of
consistent Poisson structures
for non-ultralocal classical integrable field theories.
Finally it was shown \cite{lipar} that this was the natural quadratic form
a la Sklyanin for a non-skew-symmetric $r$-matrix, reading:
\be \Big \{L_1,\ L_2 \Big \} = \Big [r- r^{\pi},\ L_1 L_2 \Big ]
+L_1 (r+r^{\pi})L_2 - L_2 (r+r^{\pi})L_1.\ee
A typical situation when one considers naturally a quadratic
Poisson structure for the Lax matrix occurs when considering
discrete (on a lattice) or continuous (on a line)
integrable systems where the Lax matrix
depends on either a discrete or a continuous variable; the Lax
pair is thus associated to a point on the space-like lattice or
continuous line \cite{ft, kulishsklyanin}. Let us first examine
the discrete case where one considers a finite set of Lax
matrices $L_n$ labelled by $n \in {\mathbb N}$.

\subsection{Periodic boundary conditions}

Introduce the Lax pair ($L,\ A$) for discrete integrable
models \cite{abla} (see also  \cite{mcwoo} for statistical
systems), and the associated auxiliary problem (see e.g.
\cite{ft}) \be  && \psi_{n+1} = L_n\ \psi_n \non\\
&& \dot{\psi}_n = A_n\ \psi_n.  \label{lat} \ee From the latter
equations one may immediately obtain the discrete zero curvature condition:
\be \dot{L}_n = A_{n+1}\ L_n - L_n\ A_n.
\label{zero}\ee The monodromy matrix arises from the first equation
(\ref{lat}) (see e.g. \cite{ft}) \be T_{a}(\lambda) = L_{aN}(\lambda)
\ldots L_{a1}(\lambda) \label{trans0}\ee where index $a$
denotes the auxiliary space, and the indices $1, \ldots , N$
denote the sites of the one dimensional classical discrete model.

Consider now a skew symmetric classical $r$-matrix which is a
solution of the classical Yang-Baxter equation \cite{skl, sts}\be
\Big [r_{12}(\lambda_1-\lambda_2),\
r_{13}(\lambda_1)+r_{23}(\lambda_2) \Big ]+ \Big [
r_{13}(\lambda_1),\ r_{23}(\lambda_2) \Big ] =0, \label{clyb} \ee
and let $L$ satisfy the associated Sklyanin bracket \be \Big \{ L_{a}(\lambda),\ L_b(\mu)
\Big \} = \Big [ r_{ab}(\lambda-\mu),\ L_{a}(\lambda) L_b(\mu)
\Big ]. \label{clalg} \ee It is then immediate that
that (\ref{trans0}) also satisfies (\ref{clalg}). Use of
the latter equation shows that the quantities $tr
T(\lambda)^n$ provide charges in involution, that is \be \Big \{
tr\ T^{n}(\lambda),\ tr\ T^m(\mu) \Big \}=0 \ee which again is trivial
by virtue of (\ref{clalg}). In the simple $sl_2$ case the only non
trivial quantity is $tr T(\lambda) = t(\lambda)$, that is the usual
``bulk'' transfer matrix.

Let us now briefly review how the classical $A$-operator and the
corresponding classical equations of motion are obtained from
the expansion of the monodromy matrix in the simple case of
periodic boundary conditions. Let us first introduce some useful
notation \be T_{a}(n,m;\lambda)= L_{an}(\lambda) L_{a
n-1}(\lambda) \ldots L_{am}(\lambda),  ~~~~n>m. \ee
To formulate $\{t(\lambda),\ L(\mu) \}$ or even better $\{\ln\ t(\lambda),\
L(\mu) \}$, given that usually $\ln\ tr T(\lambda)$ gives rise to
local integrals of motion, we first derive the quantity $\{
T_a(\lambda),\ L_{bn}(\mu) \}$.

\be \Big \{ T_a(\lambda),\
L_{bn}(\mu) \Big \} &=& T_a(N, n+1; \lambda)\ r_{ab}(\lambda-\mu)\
T_a(n, 1; \lambda)\ L_{bn}(\mu)\non\\ &-& L_{bn}(\mu)\ T_a(N, n;
\lambda)\ r_{ab}(\lambda-\mu)\ T_a(n-1,1; \lambda).
\label{basic0}\ee  Taking the trace over the auxiliary
space $a$, and then the logarithm  we conclude \be \Big \{\ln\
t(\lambda),\ L(\mu) \Big\} &=& t^{-1}(\lambda)\ \Big ( tr_a \{
T_a(N,n+1;\lambda)\ r_{ab}(\lambda-\mu)\ T_a(n,1;\lambda) \} \
L(\mu) \non\\ &-& L(\mu)\ tr_a \{ T_a(N,n;\lambda)\ r_{ab}(\lambda
-\mu)\ T_a(n-1,1;\lambda)\}  \Big) \non\\ \ee the auxiliary index
$b$ is suppressed from the latter expression for simplicity. Then
plugging the latter forms into the
classical equations of motions for all integrals of motion \be
{\dot L}_n(\mu) = \Big \{ t(\lambda),\ L_n(\mu) \Big \}
\label{eqmo1} \ee we conclude that the zero curvature condition (\ref{zero}) is realized by:
\be A_n(\lambda, \mu) =
t^{-1}(\lambda)\ tr_{a} \{ T_a(N,n;\lambda)\ r_{ab}(\lambda -\mu)\
T_a(n-1,1;\lambda) \}. \label{laf} \ee Let us focus on the classical rational
$r$-matrix (\ref{rr}). In this case $A_n$
takes the simple form: \be A_n(\lambda, \mu) = {t^{-1}(\lambda)
\over \lambda -\mu}\  T(n-1,1;\lambda)\ T(N,n;\lambda). \ee In the
open case, as we shall see in the subsequent section, it is
sufficient to consider the expansion of $t(\lambda)$ in order to
obtain the local integrals of motion.

\subsection{Open boundary conditions}

We now generalize the procedure described in the preceding
section to the case of generic integrable ``boundary conditions''.
We propose a construction of two types of monodromy and transfer
matrices, and associated Lax-type evolution equations,
of the form (\ref{clalg})--(\ref{laf}), albeit incorporating
a supplementary set of non-dynamical parameters encapsulated
into a ``reflection'' matrix $K(\lambda)$.  In some examples they may
indeed be interpreted as boundary effects consistent with integrability
of an open chain-like system. We
should stress that this is the first time to our knowledge that
such an investigation is systematically undertaken. There are related
studies regarding particular examples of open spin chains
such as XXZ, XYZ and 1D Hubbard models \cite{guan, lisa}.
However, the derivation of the corresponding Lax pair is
restricted to the Hamiltonian only and not to all associated
integrals of motion of the open chain. In this study we present a
generic description independent of the choice of model, and we
derive the Lax pair for each one of the entailed boundary
integrals of motion. Particular examples are also
presented.

The two types of monodromy matrices will respectively represent
the classical version of the reflection algebra ${\mathbb R}$,
and the twisted Yangian ${\mathbb T}$ written in the compact form:
(see e.g. \cite{sklyanin, maillet, mailfrei2}): \be && \Big \{{\cal T}_1(\lambda_1),\
{\cal T}_2(\lambda_2) \Big \} = r_{12}(\lambda_1-\lambda_2){\cal
T}_{1}(\lambda_1){\cal T}_2(\lambda_2) -{\cal T}_1(\lambda_1)
{\cal T}_2(\lambda_2) \hat r_{12}(\lambda_1 -\lambda_2) \non\\ &&
+ {\cal T}_{1}(\lambda_1) \hat r^*_{12}(\lambda_1+\lambda_2){\cal
T}_2(\lambda_2)- {\cal T}_{2}(\lambda_2)
r^*_{12}(\lambda_1+\lambda_2){\cal T}_1(\lambda_1) \label{refc}
\ee where $\hat r,\ r^*,\ \hat r^*$ are defined in (\ref{notation00}).
The latter equation may be thought of as the semiclassical
limit of the reflection equation \cite{sklyanin}. In
most well known physical cases, such as the $A^{(1)}_{{\cal N}
-1}$ $r$-matrices $r_{12}^{t_1 t_2} = r_{21}$ implying that in the
SNP case $r^*_{ab} = \hat r^*_{ab}$. In the case of the Yangian
$r$-matrix $r_{12} =r_{21}$, hence all the expressions above may
be written in a more symmetric form. These two distinct algebras
are respectively associated with the SP boundary conditions
(${\mathbb R}$) and the SNP boundary conditions (${\mathbb T}$).

In order to construct representations of (\ref{refc}) yielding
the generating function of the integrals of
motion one now introduces $c$-number representations ($K^{\pm}$) of the
algebra ${\mathbb R}$ (${\mathbb T}$) satisfying (\ref{refc}) for
SP and SNP respectively, and also the non-dynamical condition: \be \Big \{
K_1^{\pm}(\lambda_1),\ K_2^{\pm}(\lambda_2) \Big \}=0. \label{kso}
\ee
Taking now as $T(\lambda)$ any bulk monodromy matrix (\ref{trans0})
built from local $L$ matrices obeying
(\ref{clalg}) and defining in addition \be \hat T(\lambda) =
T^{-1}(-\lambda) ~~~\mbox{for SP}, ~~~~~\hat T(\lambda) =
T^{t}(-\lambda) ~~~\mbox{for SNP}.
\label{notation} \ee one gets:
\\
\\
{\it \underline{Theorem 3.1.}}: Representations of the corresponding
algebras ${\mathbb R},\
{\mathbb T}$, are given by the following expression see e.g.
\cite{sklyanin, paper}: \be && {\cal T}(\lambda) = T(\lambda)\
K^{-}(\lambda)\ \hat T(\lambda). \label{reps} \ee For a detailed
proof see e.g. \cite{paper}. $\blacksquare$

Define now as generating function of the involutive quantities \be
t(\lambda)= tr\{ K^{+}(\lambda)\ {\cal T}(\lambda)\}. \label{gen}
\ee
\\
{\it \underline{Theorem 3.2.}}: Due to (\ref{refc}) it is shown that \cite{sklyanin,
paper} \be  \Big \{t(\lambda_1),\ t(\lambda_2) \Big \} =0, ~~~
\lambda_1,\ \lambda_2 \in {\mathbb C}. \label{bint} \ee $\blacksquare$

The expansion of $t(\lambda)$ naturally gives rise to the integrals of
motions i.e. \be t(\lambda) = \sum_{i} {{\cal H}^{(i)} \over
\lambda^i}. \ee Usually one considers the quantity $\ln\ t(\lambda)$ to get
{\it local} integrals of motion, however for the examples we are
going to examine here the expansion of $t(\lambda)$ is enough to
provide the associated local quantities as will be transparent in
the subsequent section. Finally one has:
\\
\\
{\it \underline{Theorem 3.3.}}: Time evolution of the local Lax matrix $L_n$
under generating  Hamiltonian action of $t(\lambda)$ is given by:
\be {\dot L}_n(\mu) = {\mathbb A}_{n+1}(\lambda, \mu)\
L_n(\mu) - {\mathbb A}_n(\lambda, \mu)\ L_n(\mu), \ee where
${\mathbb A}_n$ is the modified (boundary) quantity,
\be {\mathbb A}_n(\lambda, \mu) &=& tr_a \Big (K_a^+(\lambda)\
T_a(N, n;\lambda)\ r_{ab}(\lambda-\mu)\ T_a(n-1, 1;\lambda)\
K^-_a(\lambda)\ \hat T_a(\lambda) \non\\ &+& K_a^+(\lambda)\
T_a(\lambda)\ K^-_a(\lambda)\ \hat T_a(1,n-1;\lambda)\ \hat
r_{ab}^*(\lambda +\mu)\ \hat T_a(n, N;\lambda) \Big ).
\label{aan1} \ee
\\
{\it \underline{Proof}}: We need
in addition to (\ref{clalg}) one more fundamental relation i.e.
\be \Big \{\hat L_{a}(\lambda), L_b(\mu) \Big \} = \hat
L_a(\lambda) \hat r^*_{ab}(\lambda)L_b(\mu) - L_b(\mu) \hat
r^*_{ab}(\lambda+\mu) \hat L_a(\lambda). \label{funda2} \ee 
where the notation $\hat L$ is self-explanatory from (\ref{notation}).
Taking into account the latter expressions we derive for $\hat T$
\be \Big \{ \hat T_a(\lambda),\ L_{bn}(\mu) \Big \} &=& \hat
T_a(1,n; \lambda)\ \hat r^*_{ab}(\lambda+\mu)\ \hat T_a(n+1, N;
\lambda)\ L_{bn}(\mu)\non\\ &-& L_{bn}(\mu)\ \hat T_a(1, n-1;
\lambda)\ \hat r^*_{ab}(\lambda+\mu)\ \hat T_a(n, N; \lambda).
\label{basic2} \ee The next step is to formulate $\Big \{
t(\lambda),\ L_{bn}(\mu) \Big \}$; indeed recalling
(\ref{basic0}), (\ref{basic2}), (\ref{reps}) and (\ref{gen}) we
conclude: \be
\Big \{t(\lambda),\ L_{bn}(\mu) \Big\}&=& tr_a \Big
(K_a^+(\lambda)\ T_a(N, n+1;\lambda)\  r_{ab}(\lambda-\mu)\ T_a(n,
1;\lambda)\ K^-_a(\lambda)\ \hat T(\lambda) \non\\ &+&
K_a^+(\lambda)\ T_a(\lambda)\ K^-_a(\lambda) \hat
T_a(1,n;\lambda)\ \hat r_{ab}^*(\lambda +\mu)\ \hat T_a(n+1,
N;\lambda) \Big  ) L_{bn}(\mu) \non\\ &-& L_{bn}(\mu)\ tr_a \Big (
K_a^+(\lambda)\ T_a(N,n;\lambda)\ r_{ab}(\lambda -\mu)\
T_a(n-1,1;\lambda)\ K^-_a(\lambda) \hat T(\lambda) \non\\ &+&
K_a^+(\lambda)\ T_a(\lambda)\ K_a^-(\lambda)\ \hat
T_a(1,n-1;\lambda)\ \hat r_{ab}^*(\lambda+\mu)\ \hat T_{a}(n,
N;\lambda) \Big ). \label{modan} \ee Expression (\ref{aan1})
is readily extracted from (\ref{modan}). $\blacksquare$

Special care should be taken at the boundary
points $n=1$ and $n=N+1$. Indeed going back to formula
(\ref{aan1}) restricting ourselves to $n=1$ and $n=N+1$ and taking
into account that $T(N, N+1, \lambda)= T(0,1,\lambda) = \hat
T(1,0,\lambda) =\hat T(N+1, N, \lambda) ={\mathbb I}$ we obtain
the explicit form for ${\mathbb A}_1,\ {\mathbb A}_{n+1}$.
We should stress that the derivation of the boundary Lax pair is universal,
namely the expressions (\ref{aan1}) are generic and independent
of the choice of $L,\ r$.
Note that expansion of the latter expressions of ${\mathbb A}_n
(\lambda, \mu) = \sum{{\mathbb A}_n^{(i)} \over \lambda^i}$
provides all the ${\mathbb A}_n^{(i)}$ associated to the
corresponding integrals of motion ${\cal H}^{(i)}$, which follow
from the expansion of $t(\lambda)$. This will become quite
transparent in the examples presented in the subsequent section.

{\bf Remark}: A different construction of (\ref{refc}) was already given
in a very general setting in \cite{mailfrei2}. It is related to the formulation of non-ultralocal
integrable field theories on a lattice and extends the analysis of \cite{maillet}.
The essential difference with our construction is that the $k$ matrix (denoted $\gamma$)
in \cite{mailfrei2}, is sandwiched between the ``local'' monodromy matrices
$T_{n,n-1}$ so as to obtain an overall Poisson bracket of the form (\ref{refc})
for the ``dressed'' monodromy matrix ...$T_{n+1,n} \gamma T_{n, n-1} \gamma$....
The matrix $\gamma$ allows to take into account the effects of the non-ultralocal
part $\delta' (x-y)$ of the Poisson bracket structure. In this respect $\gamma$
also corresponds to ``boundary'' effects, although multiple and internal.

\subsection{Examples}

We shall now examine a simple example, i.e. the open
generalized DST model, which may be seen as a lattice version of
the generalized (vector) NLS model, (see also
\cite{kundubas, nls1, nls2, ragn, paper} for further details). 
The open Toda chain will be also
discussed as a limit of the DST model. We shall explicitly
evaluate the ``boundary'' Lax pairs for the first integrals of
motion. If we focus on the special case (\ref{rr}), which will be
our main interest here, the latter expression reduces to the
following expressions for SP and SNP boundary conditions. In
particular, ${\mathbb A}_n$ for SP boundary conditions reads: \be
{\mathbb A}_n(\lambda, \mu) &=& {1\over \lambda -\mu}\
T(n-1,1;\lambda)\ K^{-}(\lambda)\ \hat T(\lambda)\ K^+(\lambda)\
T(N,n;\lambda) \non\\ &+& {1\over \lambda +\mu}\ \hat T(n,
N;\lambda)\ K_a^{+}(\lambda)\ T(\lambda)\ K^{-}(\lambda)\ \hat
T(1,n-1 ; \lambda) ~~~\mbox{for SP} \label{ann}\ee and for SNP
boundary conditions: \be  {\mathbb A}_n(\lambda, \mu) &=& {1 \over
\lambda -\mu}\ T(n-1,1;\lambda)\ K^{-}(\lambda)\ \hat T(\lambda)\
K^+(\lambda)\ T(N,n;\lambda) \non\\ &-& {1 \over \lambda +\mu}\
\hat T^t(1,n-1 ; \lambda)\ K_a^{-t}(\lambda)\ T^t(\lambda)\
K^{+t}(\lambda)\ \hat T^t(n, N;\lambda) ~~~\mbox{for SNP}, \non\\
\label{a1N} \ee where we recall that for the special points $n=1,\
N+1$ we take into account that: $T(N, N+1, \lambda)=
T(0,1,\lambda) = \hat T(1,0,\lambda) =\hat T(N+1, N, \lambda)
={\mathbb I}$.

The Lax operator of the $gl({\cal N})$ DST model has the following
form: \be L(\lambda) = (\lambda - \sum_{j=1}^{{\cal N}-1} x^{(j)}
X^{(j)}) E_{11} +b\sum_{j=2}^{{\cal N}} E_{jj} +b\sum_{j=2}^{{\cal
N}}x^{(j-1)} E_{1j} - \sum_{j=2}^{{\cal N}}X^{(j-1)}E_{j1}  \ee
with $x^{(j)}_n,\ X^{(j)}_n$ being canonical variables \be && \Big \{
x^{(i)}_{n},\ x^{(j)}_m \Big \}= \Big \{ X_{n}^{(i)},\ X_m^{(j)}
\Big \},~~~~~ \Big \{ x_n^{(i)},\ X_m^{(j)} \Big \} = \delta_{nm}
\delta_{ij} \non\\ && i,\ j \in \{1, \ldots, {\cal N} \}, ~~~n,\ m
\in \{1, \ldots, N \}. \label{canon} \ee In \cite{paper} the first
non-trivial integral of motion for the SNP case, choosing the simplest
consistent value $K^{\pm} = {\mathbb I}$ was explicitly computed:
\be && {\cal H} = -{1\over 2} \sum_{n=1}^N {\mathbb
N}_n^2 -b \sum_{n=1}^N \sum_{j=1}^{{\cal N}-1}
X_n^{(j)}x_{n+1}^{(j)} - {1\over 2}\sum_{j=1}^{{\cal
N}-1}(X_N^{(j)}X_N^{(j)} +b^2x_1^{(j)} ) \non\\ && \mbox{where}
~~~{\mathbb N}_n = \sum_{j=1}^{{\cal N}-1}x_n^{(j)}X_{n}^{(j)}.
\label{ham0}\ee Our aim is now to determine the modified Lax
pair induced by the non-trivial integrable boundary conditions. We
shall focus here on the case of SNP boundary conditions, basically
because in the particular example we consider here such boundary
conditions are technically easier to study. Moreover, the SNP
boundary conditions have not been so much analyzed in the context of lattice
integrable models, which provides an extra motivation to
investigate them. Taking into account (\ref{a1N}) we
explicitly derive the modified Lax pair for the generalized
DST model with SNP boundary conditions. Indeed, after expanding
(\ref{a1N}) in powers of $\lambda^{-1}$ we obtain the quantity
associated to the Hamiltonian (\ref{ham0})\footnote{To obtain both
${\cal H}^{(2)}$ and ${\mathbb A}^{(2)}_n$ we divided the original
expressions by a factor two.}: \be && {\mathbb A}_n^{(2)} = \lambda E_{11} -
\sum_{j\neq 1} X_{n-1}^{(j-1)}E_{j1}  +b \sum_{j\neq 1}
x_{n}^{(j-1)}E_{1j}, ~~~n \in \{ 2, \ldots N\} \non\\ &&  {\mathbb
A}_1^{(2)}= \lambda E_{11} - b \sum_{j \neq 1} x_{1}^{(j-1)}E_{j1}
+b \sum_{j\neq 1}  x_{1}^{(j-1)}E_{1j}, \non\\ &&  {\mathbb
A}_{N+1}^{(2)} = \lambda E_{11} -  \sum_{j \neq 1}
X_{N}^{(j-1)}E_{j1}  + \sum_{j\neq 1}  X_{N}^{(j-1)}E_{1j}.
\label{aa} \ee Let us now consider the simplest possible case,
i.e. the $sl_2$ DST model. It is worth stressing that in this case
the SP and SNP boundary coincide given that \be L^{-1}(-\lambda) =
V\ L^t(-\lambda)\ V, ~~~~V =\mbox{antid}(1, \ldots,1).
\label{gag}\ee In this particular case the Hamiltonian
(\ref{ham0}) reduces into:  \be {\cal H}^{(2)} = -{1\over 2}
\sum_{n=1}^N x_{n}^2 X_n^2 -b \sum_{n=1}^{N-1} x_{n+1} X_n -{b^2
\over 2} x_1^2 -{1 \over 2} X_N^{2}. \label{hami}\ee The equations
of motion associated to the latter Hamiltonian may be readily
extracted by virtue of\footnote{The associated Poisson brackets
for both DST and Toda models are defined as: \be && \Big \{A,\ B
\Big \} = \sum_n\ \Big ({\partial A \over \partial x_n}\ {\partial
B \over \partial X_n}- {\partial A \over \partial X_n}\ {\partial
B \over \partial x_n} \Big ), ~~~\mbox{DST model} \non\\ && \Big
\{A,\ B \Big \} = \sum_n\ \Big ({\partial A \over \partial q_n}\
{\partial B \over
\partial p_n}- {\partial A \over \partial p_n}\ {\partial B \over
\partial q_n} \Big ), ~~~\mbox{Toda chain}. \ee } \be {\dot L}
= \Big \{ {\cal H}^{(2)},\ L \Big \}. \label{com}\ee
It is deduced from (\ref{aa}) for the $sl_2$ case: \be &&
{\mathbb A}_n^{(2)}(\lambda) =  \left(
\begin{array}{cc}
\lambda  &b x_n \\
-X_{n-1}   &0
\end{array} \right ), ~~~~n \in \{2, \ldots, N \} \non\\
&& {\mathbb A}_1^{(2)}(\lambda) =  \left(
\begin{array}{cc}
\lambda  &b x_1 \\
-bx_{1}   &0
\end{array} \right ), ~~~~ {\mathbb A}_{N+1}^{(2)}(\lambda) =  \left(
\begin{array}{cc}
\lambda  &X_N \\
-X_{N}   &0
\end{array} \right ). \ee Alternatively the equations of motion may be
derived from the zero curvature condition \be {\partial L_n \over
\partial t} = {\mathbb A}_{n+1}^{(i)}\ L_n - L_n {\mathbb
A}_{n}^{(i)} \label{zerob} \ee which the modified Lax pair
satisfies. It is clear that to each one of the higher local
charges a different quantify ${\mathbb A}_{n}^{(i)}$ is associated. Both
equations (\ref{com}), (\ref{zerob}) lead naturally to the same
equations of motion, which for this particular example read as: \be
&& {\dot x}_n = x_n^2 X_n +b  x_{n+1}, ~~~~{\dot X}_n = -x_n
X_n^{2} -b X_{n-1}, ~~~n\in \{2, \ldots N-1 \} \non\\  && {\dot
x}_1 = x_1^2 X_1 + b x_2, ~~~~{\dot X}_1 =  -x_1 X_1^2 - b x_1
\non\\ && {\dot x}_N = x_N^2 X_N +X_N, ~~~~ {\dot X}_N = -x_N
X_N^2 - bX_{N-1}. \ee

The Toda model may be seen as an appropriate limit of the DST
model (see also \cite{skld}). Indeed consider the following
limiting process as $b \to 0$: \be X_n \to e^{-q_n}, ~~~~x_n \to
e^{q_n} (b^{-1} +p_n)\ee It is clear that the harmonic oscillator
algebra defined by $(x_n,\ X_n,\ x_n X_n)$ reduces to the
Euclidian Lie algebra $(e^{\pm q_n}, \ p_n)$, and consequently the
Lax operator reduces to: \be L_n(\lambda)= \left(
\begin{array}{cc}
\lambda -p_n   &e^{q_n} \\
 -e^{-q_n}     &0
\end{array} \right )\ee  where $q_n,\ p_n$ are canonical variables. In this
case the corresponding Hamiltonian may be readily extracted and
takes the form \be {\cal H}^{(2)} = -{1\over 2} \sum_{n=1}^N p_n^2 -
\sum_{n=1}^{N-1} e^{q_{n+1} - q_n} -{1\over 2} e^{2q_1} -{1 \over
2} e^{-2q_{N}} \ee and the corresponding ${\mathbb A}_n^{(2)}$ are
expressed as \be && {\mathbb A}_n^{(2)}(\lambda) = \left(
\begin{array}{cc}
\lambda  & e^{q_n} \\
-e^{q_{n-1}}   &0
\end{array} \right ), ~~~~n \in \{2, \ldots, N \} \non\\ &&
{\mathbb A}_1^{(2)}(\lambda) =  \left(
\begin{array}{cc}
\lambda  &e^{q_1} \\
-e^{q_1}   &0
\end{array} \right ), ~~~~ {\mathbb A}_{N+1}^{(2)}(\lambda) =  \left(
\begin{array}{cc}
\lambda  &e^{-q_N} \\
-e^{-q_N}   &0
\end{array} \right ). \ee
In this case as well both formulas (\ref{zerob}) and (\ref{com})
lead to the following set of equations of motions: \be && p_n=
{\dot q}_n, ~~~~{\ddot q}_n = e^{q_{n+1} -q_n}- e^{q_n - q_{n-1}},
~~~~n \in \{2, \ldots, N-1 \} \non\\ && p_1 = {\dot q}_1,
~~~~{\ddot q}_1 =e^{q_2 -q_1} - e^{2q_1} \non\\ && p_N = {\dot
q}_N, ~~~~{\ddot q}_N = e^{-2q_N} - q^{q_N -q_{N-1}}. \ee

\section{The continuous case}

\subsection{Periodic boundary conditions}

Let us now recall the basic notions regarding the Lax pair and the
zero curvature condition for a continuous integrable model following
essentially \cite{ft}. Define $\Psi$ as being a solution of the following set
of equations (see e.g. \cite{ft}) \be &&{\partial \Psi \over \partial
x} = {\mathbb U}(x,t, \lambda) \Psi \label{dif1}\\ && {\partial \Psi
\over \partial t } = {\mathbb V}(x,t,\lambda)\Psi \label{dif2} \ee 
${\mathbb U},\ {\mathbb V}$ being in general $n \times n$ matrices with entries
defined as functions of complex valued dynamical fields, their derivatives,
and the spectral parameter $\lambda$. 
Compatibility conditions of the two differential
equation (\ref{dif1}), (\ref{dif2}) lead to the zero curvature
condition \cite{AKNS}--\cite{ZSh} \be \dot{{\mathbb U}} - {\mathbb V}' + \Big [{\mathbb
U},\ {\mathbb V} \Big ]=0, \label{zecu} \ee giving rise to the
corresponding classical equations of motion of the system under
consideration. The monodromy matrix may be written from (\ref{dif1}) as: 
\be T(x,y,\lambda) = {\cal P} exp \Big \{
\int_{y}^x {\mathbb U}(x',t,\lambda)dx' \Big \}, \label{trans} \ee 
where apparently $T(x, x, \lambda) =1$.
The fact that the monodromy matrix satisfies equation (\ref{dif1}) is
extensively used to get the relevant integrals
of motion and the associated Lax pairs.

Hamiltonian formulation of the equations of motion is available again under
the $r$-matrix approach. In this picture the underlying
classical algebra is manifestly analogous to the quantum case. Let us first recall this
method for a general classical integrable system on the full line.
The existence of the Poisson structure for ${\mathbb U}$ realized by the classical r-matrix,
satisfying the classical Yang-Baxter equation (\ref{clyb}), guarantees the integrability of
the classical system. Indeed assuming that the operator ${\mathbb U}$ satisfies the following
ultralocal form of Poisson brackets
\be \Big \{{\mathbb U}_a(x, \lambda),\ {\mathbb U}_b(y, \mu) \Big \} =
\Big [r_{ab}(\lambda - \mu),\ {\mathbb U}_a(x, \lambda) +{\mathbb U}_b (y,\mu) \Big ]\
\delta(x-y), \label{ff0} \ee
$T(x,y,\lambda)$
satisfies: \be \Big \{T_{a}(x,y,t,\lambda_1),\
T_{b}(x,y,t,\lambda_2) \Big \} =
\Big[r_{ab}(\lambda_1-\lambda_2),\
T_a(x,y,t,\lambda_1)T_b(x,y,t,\lambda_2) \Big ]. \label{basic} \ee
Making use of the latter equation one may readily show for a
system on the full line: \be \Big \{\ln tr\{T(x,y,\lambda_1)\},\ \ln
tr\{T(x,y, \lambda_2)\} \Big\}=0 \ee i.e. the system is
integrable, and the charges in involution --local integrals of
motion-- are obtained by expansion of the generating function $\ln
tr\{T(x,y,\lambda)\}$, based essentially on the fact that $T$
satisfies (\ref{dif1}).

Let us now recall how one constructs the ${\mathbb V}$-operator associated to
given local integrals of motion. One easily proves the following
identity using (\ref{ff0}) \be  \Big \{ T_a(L,
-L, \lambda),\  {\mathbb U}_b(x, \mu) \Big \}= {\partial M(x,
\lambda, \mu) \over \partial x} + \Big [ M(x, L, -L, \lambda,
\mu),\ {\mathbb U}_{b}(x, \mu) \Big ] \label{basic1} \ee where we
define \be M(x, \lambda, \mu) = T_{a}(L, x, \lambda)\
r_{ab}(\lambda -\mu)\ T_a(x, L, \lambda). \label{first} \ee For
more details on the proof of the formula above we refer the
interested reader to \cite{ft}; (\ref{basic1}) may be
seen as the continuum version of relation (\ref{basic}). Recall that 
$t(\lambda)= tr T(\lambda)$ then it
naturally follows from (\ref{basic1}), and (\ref{zecu}), that
\be \Big \{ \ln\ t(\lambda),\ {\mathbb U}(x,
\lambda) \Big \} = {\partial {\mathbb V}(x, \lambda, \mu) \over
\partial x} + \Big [ {\mathbb V}(x, \lambda, \mu),\ {\mathbb
U}(x,\lambda) \Big ] \ee with \be {\mathbb V}(x, \lambda, \mu) =
t^{-1}(\lambda)\ tr_a \Big ( T_a(L, x, \lambda)\ r_{ab}(\lambda,
\mu)\ T_a( x, -L, \lambda) \Big ) \label{vv} \ee and in the rational
case (\ref{vv}) reduces to \be {\mathbb V}(x, \lambda, \mu)=
{t^{-1}(\lambda) \over \lambda -\mu}\ T(x, -L, \lambda)\ T(L, x,
\lambda). \ee

\subsection{General integrable boundary conditions}

Our aim here is to consider integrable models on the interval
with consistent ``boundary conditions'', and
derive rigorously the Lax pairs associated to the entailed
boundary local integrals of motion as a continuous extension of
theorems 3.1.--3.3. For this purpose we follow the
line of action described in \cite{ft}, using now Sklyanin's
formulation for the system on the interval or on the half line. We
briefly describe this process below for any classical integrable
system on the interval. In this case one constructs a modified
`monodromy' matrix ${\cal T}$, based on Sklyanin's formulation and
satisfying again the Poisson bracket algebras ${\mathbb R}$ or
$~{\mathbb T}$. To
construct the generating function of the integrals of motion one
also needs $c$-number representations of the algebra ${\mathbb R}$
or ${\mathbb T}$ satisfying (\ref{refc}) for SP and SNP
respectively, such that: \be \Big \{ K_1^{\pm}(\lambda_1),\
K_2^{\pm}(\lambda_2) \Big \}=0. \label{kso1} \ee
\\
{\it \underline{Theorem 4.1.}}:
The modified `monodromy' matrices, realizing the corresponding algebras
${\mathbb R},\ {\mathbb T}$, are given by the following expressions
\cite{sklyanin} ($\hat T$ being defined in (\ref{notation})):
\be  && {\cal T}(x,y,t,\lambda) =
T(x,y,t,\lambda)\ K^{-}(\lambda)\ \hat T(x,y,t,\lambda).
\label{reps0} \ee $\blacksquare$

The generating function of the involutive
quantities is defined as \be t(x,y,t,\lambda)= tr\{
K^{+}(\lambda)\ {\cal T}(x,y,t,\lambda)\}. \ee
Indeed one shows:
\\
{\it \underline{Theorem 4.2.}}:
\be  \Big \{t(x,y,t,\lambda_1),\
t(x,y,t,\lambda_2) \Big \} =0, ~~~ \lambda_1,\ \lambda_2 \in
{\mathbb C}. \label{bint0} \ee $\blacksquare$

In the case of open boundary conditions, exactly as in the discrete integrable models,
and taking into account (\ref{basic2}) we prove \be \Big \{{\cal
T}_a(0, -L,\lambda),\ {\mathbb U}_b(x, \mu) \Big \} = {\mathbb M}_a'(x,
\lambda, \mu) + \Big [{\mathbb M}_a(x, \lambda, \mu),\ {\mathbb
U}_b(x,\ \mu ) \Big] \label{last} \ee where we define \be {\mathbb
M}(x,\lambda, \mu) &=& T(0, x, \lambda) r_{ab}(\lambda -\mu) T(x,
-L, \lambda) K^-(\lambda) \hat T(0, -L, \lambda) \non\\ &+& T(0,
-L, \lambda) K^-(\lambda) \hat T(x, -L, \lambda) \hat r^*_{ab}(\lambda
+\mu) \hat T(0, x, \lambda). \label{mm} \ee Finally bearing in mind the
definition of $t(\lambda)$ and (\ref{last}) we conclude with:
\\
\\
{\it \underline{Theorem 4.3.}}: \be \Big
\{ \ln\ t(\lambda),\ {\mathbb U}(x, \mu) \Big \} = {\partial {\mathbb
V}(x,\lambda, \mu) \over \partial x} + \Big [ {\mathbb V}(x,\lambda, \mu),\
{\mathbb U}(x, \mu) \Big ]
\label{defin} \ee where \be {\mathbb V}(x,\lambda,
\mu) = t^{-1}(\lambda) \ tr_a \Big ( K^+(\lambda)\ {\mathbb M}_a(x , \lambda, \mu) \Big ).
\label{final1} \ee $\blacksquare$

As in the discrete case particular
attention should be paid to the boundary points $x =0,\ -L$. Indeed, for these
two points one has to simply take into account that $T(x,x
,\lambda) = \hat T(x,x, \lambda) ={\mathbb I}$. Moreover, the expressions derived in
(\ref{mm}), (\ref{final1})
are universal, that is independent of the choice of model. As was remarked upon
when discussing the discrete case,
a quadratic algebra of the form (\ref{refc}) was initially obtained in
the continuous case \cite{maillet} when extending the derivation of 4.1 to situations
where the Poisson brackets (\ref{ff0}) are non-ultralocal, exhibiting $\delta'(x-y)$ terms.
Connection to ``boundary'' effects was discussed previously (see Section 3.2).

\subsection{Example}

We shall now examine a particular example associated to the
rational $r$-matrix (\ref{rr}), that is the $gl_{{\cal N}}$ NLS
model. Although in \cite{paper} an extensive analysis for both
types of boundary conditions is presented, here we shall focus on
the simplest diagonal ($K^{\pm} ={\mathbb I}$) boundary
conditions. The Lax pair is given by the following
expressions \cite{ft, foku}: \be {\mathbb U} = {\mathbb U}_0 +
\lambda {\mathbb U}_1, ~~~{\mathbb V} = {\mathbb
V}_0+\lambda{\mathbb V}_1 +\lambda^2 {\mathbb V}_2 \label{lax0}\ee
where \be && {\mathbb U}_1 = {1\over 2i} (\sum_{i=1}^{{\cal
N}-1}E_{ii} -E_{{\cal N}{\cal N}}), ~~~~{\mathbb U}_0 =
\sum_{i=1}^{{\cal N}-1}(\bar \psi_i E_{i{\cal N}} +\psi_i E_{{\cal N}i}) \non\\
&&{\mathbb V}_0 = i \sum_{i,\ j=1}^{{\cal N}-1}(\bar \psi_i \psi_j E_{ij}
-|\psi_i|^2E_{{\cal N}{\cal N}}) -i\sum_{i=1}^{{\cal N}-1} (\bar \psi_i' E_{i{\cal N}} -
\psi_i' E_{{\cal N}i}), \non\\ && {\mathbb V}_1= -{\mathbb U}_{0},
~~~{\mathbb V}_2= -{\mathbb U}_{1} \label{lax} \ee and $\psi_i,\
\bar \psi_j$ satisfy\footnote{The Poisson structure for the
generalized NLS model is defined as: \be \Big \{ A,\ B  \Big \}= i
\sum_{i} \int_{-L}^{L} dx \Big ({\delta A \over \delta \psi_i(x)}\
{\delta B \over \delta \bar \psi_i(x)} - {\delta A \over \delta
\bar \psi_i(x)}\ {\delta B \over \delta \psi_i(x)}\Big )  \ee}:
\be \Big \{ \psi_{i}(x),\ \psi_j(y) \Big \} = \Big \{\bar
\psi_{i}(x),\ \bar \psi_j(y) \Big \} =0, ~~~~\Big \{\psi_{i}(x),\
\bar \psi_j(y) \Big \}= \delta_{ij}\ \delta(x-y). \ee Note that we
have suppressed the constant $\kappa$ from (\ref{lax}) compared
e.g. to \cite{paper} by rescaling the fields
$(\psi_i,\ \bar \psi_i ) \to \sqrt{\kappa} (\psi_i,\ \bar \psi_i)$.

From the zero curvature condition (\ref{zecu}) the classical
equations of motion for the generalized NLS model with periodic boundary conditions
are entailed i.e. \be i{\partial \psi_{i}(x,t) \over
\partial t} = - {\partial^{2} \psi_{i}(x,t) \over
\partial^2 x}+2\kappa \sum_{j}|\psi_{j}(x,t)|^2 \psi_{i}(x,t),
~~~~i,\ j \in \{1, \ldots, {\cal N}-1 \}. \label{nls} \ee It is clear
that for ${\cal N}=2$ the equations of motion of the usual NLS model are
recovered. The boundary Hamiltonian for the generalized NLS model
may be expressed as \be {\cal H}&=& \int_{-L}^0 dx\
\sum_{i=1}^{{\cal N}-1}\Big (\kappa |\psi_{i}(x)|^2
\sum_{j=1}^{{\cal N}-1}|\psi_j(x)|^2 +\psi'_i(x) \bar \psi'_i(x)\Big )
\non\\ &-& \sum_{i=1}^{{\cal N}-1} \Big (\psi'_i(0) \bar \psi_i(0)
+ \psi_i(0) \bar \psi'_i(0)\Big ) + \sum_{i=1}^{{\cal N}-1}  \Big
(\psi_i'(-L) \bar \psi_i(-L) + \psi_i(-L) \bar \psi'_i(-L)\Big ).
\ee One sees here that the $K$-matrix indeed contributes as a genuine boundary effect.
The Hamiltonian, obtained as one of the charges in
involution (see e.g. \cite{paper} for further details) provides the
classical equations of motion by virtue of: \be && {\partial
\psi_i(x,t)\over
\partial t} = \Big \{{\cal H}(0,-L),\ \psi_i(x,t) \Big \}, ~~{\partial
\bar \psi_i(x,t)\over \partial t} = \Big \{{\cal H}(0,-L),\ \bar
\psi_i(x,t) \Big \}, \non\\ && -L \leq x \leq 0. \label{eqmo0} \ee
Indeed considering the Hamiltonian ${\cal H}$, we end up with the
following set of equations with Dirichlet type boundary conditions\be && i {\partial \psi_i(x,t) \over
\partial t} = - {\partial^2 \psi_i(x,t)\over  \partial^2 x} +2\kappa
\sum_{j=1}^{{\cal N}-1}|\psi_{j}(x,t)|^2\psi_i(x,t) \non\\ && \psi_i(0) = \psi_i(-L) =0 ~~~~~i
\in \{1, \ldots ,{\cal N}-1\}.  \label{eqmo2} \ee For a detailed and quite exhaustive analysis
of the various integrable boundary conditions of the NLS model see \cite{paper}.
Note also that the ${\cal N}=2$ case was investigated
classically on the half line in \cite{tarasov}, whereas the NLS
equation on the interval was studied in \cite{fokas}.

As mentioned our ultimate goal here is to derive the boundary Lax
pair, in particular the ${\mathbb V}$ operator. In general for any $gl_{\cal N}$ $r$-matrix we may express
(\ref{final1}), taking also into account (\ref{defin}), for the
two types of boundary conditions already described in the first
section, i.e. \be {\mathbb V}(x, \lambda, \mu) &=&
{t^{-1}(\lambda) \over \lambda -\mu} T(x, -L, \lambda)
K^-(\lambda) \hat T(0, -L, \lambda) K^+(\lambda) T(0 ,x, \lambda)
\non\\ &+&  {t^{-1}(\lambda) \over \lambda + \mu} \hat T(0, x,
\lambda) K^+(\lambda) \ T(0, -L, \lambda) K^-(\lambda) \hat T(x,
-L, \lambda) ~~~\mbox{for SP} \label{sp} \ee and for the SNP
boundary conditions we obtain: \be {\mathbb V}(x, \lambda, \mu)
&=& {t^{-1}(\lambda) \over \lambda -\mu} T(x, -L, \lambda)
K^-(\lambda) \hat T(0, -L, \lambda) K^+(\lambda) T(0 ,x, \lambda)
\non\\ &+&  {t^{-1}(\lambda) \over \lambda + \mu} \hat T^t(x, -L,
\lambda) K^{-t}(\lambda)\ T^t(0, -L, \lambda) K^{+t}(\lambda) \hat
T^t(0, x, \lambda) ~~~\mbox{for SNP} \label{snp} \ee Again for the
boundary points $x =0,\ -L$ we should bear in mind that $T(x,x,
\lambda) =\hat T(x, x, \lambda) = {\mathbb I}$. Ultimately we wish
to expand $T(\lambda),\ \hat T(\lambda)$ in powers of
$\lambda^{-1}$ in order to determine the Lax pair for each one of
the integrals of motion. For a detailed description of the
derivation of the boundary integrals of motion for the generalized
NLS models see \cite{paper}. Hereafter we shall focus on the SP
case with the simplest boundary conditions i.e. $K^{\pm} ={\mathbb
I}$. Expanding the expression (\ref{sp}) in powers of
$\lambda^{-1}$ (we refer the interested reader to Appendix C
for technical details) we conclude that ${\mathbb
V}^{(3)}(x,\lambda)$ --the bulk part-- coincides with ${\mathbb
V}$ defined in (\ref{lax0}), (\ref{lax}), and for the boundary
points $x_b \in \{0,\ -L\}$ in particular: \be {\mathbb
V}^{(3)}(x_b, \lambda) = -{\lambda^2 \over 2i} \Big
(\sum_{i=1}^{{\cal N}-1} E_{ii} -E_{{\cal N}{\cal N}} \Big ) + i \sum_{i, j =1} ^{{\cal N}-1}
\bar \psi_i(x_b) \psi_j(x_b) E_{ij}  -i \sum_{i,j =1}^{{\cal N}-1} \Big (
\bar \psi'_i(x_b) E_{i{\cal N}} -\psi'_i(x_b) E_{{\cal N}i} \Big). \non\\ \ee We
may alternatively rewrite the latter formula as: \be {\mathbb
V}^{(3)}(x_b, \lambda) = {\mathbb V}(x_b, \lambda) +i
\sum_{i=1}^{{\cal N}-1} |\psi_i(x_b)|^2 E_{{\cal N}{\cal N}} + \lambda \sum_{i=1}^{{\cal N}-1}
(\bar \psi_i(x_b) E_{i{\cal N}} + \psi_i(x_b) E_{{\cal N}i}). \ee The last two
terms additional to ${\mathbb V}$ (\ref{lax0}), (\ref{lax}) are
due to the non-trivial boundary conditions; of course more
complicated boundary conditions would lead to more intricate
modifications of the Lax pair ${\mathbb V}$, however such an
exhaustive analysis is beyond the intended scope of the present
investigation. It can be shown that the
modified Lax pair $({\mathbb U},\ {\mathbb V}^{(3)})$ gives rise
to the classical equations of motion (\ref{eqmo2}). It is clear
that the `bulk' quantity ${\mathbb V}$ in the case of SP boundary
conditions remains intact. In the SNP case on the other hand we
may see that even the bulk part of the Lax pair is drastically
modified, due to the fact that the bulk part of the corresponding
integrals of motions is also dramatically altered. We shall not further
comment on this point, which will be anyway treated in
full detail elsewhere.

\noindent{\bf Acknowledgments:} This work was supported by
INFN, Bologna section, through grant TO12.
J.A. wishes to thank INFN and University of Bologna for hospitality.

\appendix

\section{Appendix}

We present here technical details on the derivation of the
conserved quantities for the generalized NLS model on the full
line (see also \cite{paper}). Recall that for $\lambda \to \pm i\infty $ one may express
$T$ as \cite{ft} \be T(x,y,\lambda) = ({\mathbb I} +W(x,
\lambda))\ \exp[Z(x,y,\lambda)]\ ({\mathbb I} +W(y,\lambda))^{-1}
\label{exp0} \ee where $W$ is an off diagonal matrix i.e. $~W =
\sum_{i\neq j} W_{ij} E_{ij}$, and $Z$ is purely diagonal $~Z =
\sum_{i=1}^{\cal N} Z_{ii}E_{ii}$. Also \be Z_{ii}(\lambda) =
\sum_{n=-1}^{\infty} {Z^{(n)}_{ii} \over \lambda^{n}}, ~~~~W_{ij}
= \sum_{n=1}^{\infty}{W_{ij}^{(n)} \over \lambda^n}. \label{expa}
\ee The first step is to insert the ansatz (\ref{exp0}) in
equation (\ref{dif1}).  Then we separate the diagonal and off
diagonal part and obtain the following expressions: \be && Z' =
\lambda {\mathbb U}_1 +({\mathbb U}_{0}W)^{(D)} \non\\ && W' +WZ'
= {\mathbb U}_{0} +({\mathbb U}_0 W)^{(O)} +\lambda{\mathbb U}_1 W
\label{DO} \ee where the superscripts $(D),\ (O)$ denote the
diagonal and off diagonal part of the product ${\mathbb U}_0W$.
Recall that $W = \sum_{i\neq j} W_{ij} E_{ij}, ~~Z= \sum_{i}Z_{ii}
E_{ii}$ then it is straightforward to obtain: \be && ({\mathbb
U}_0W)^{(D)} = \sqrt{\kappa} \sum_{i=1}^{{\cal N}-1}\Big ( \bar \psi_i
W_{{\cal N}i} E_{ii} + \psi_i W_{i{\cal N}} E_{{\cal N}{\cal N}} \Big ) \non\\ && ({\mathbb
U}_0W)^{(O)} = \sqrt{\kappa} \sum_{i\neq j,\ i\neq {\cal N},\ j\neq {\cal N}}
\Big (\bar \psi_i W_{{\cal N}j} E_{ij} + \psi_{i} W_{ij}E_{{\cal N}j} \Big ).
\label{contr} \ee Substituting the latter expressions
(\ref{contr}) in (\ref{DO}), we obtain \be Z(L, -L, \lambda) =
-i\lambda L \Big (\sum_{i=1}^{{\cal N}-1}E_{ii} -E_{{\cal N}{\cal N}} \Big )
+\sqrt{\kappa} \sum_{i=1}^{{\cal N}-1}\int_{-L}^{L}\ dx \Big (\bar \psi_i
W_{{\cal N}i} E_{ii} +\psi_i W_{i{\cal N}} E_{{\cal N}{\cal N}}\Big ). \ee
The leading contribution in the expansion of ($\ln\ tr T$), (where $T$
is given in (\ref{exp0})) for $i\lambda \to \infty $ comes
from $Z_{{\cal N}{\cal N}}$, with a leading term $i\lambda L$
(all other $Z_{ii},\ i\neq {\cal N}$ have a $-i \lambda L$
leading term, so when exponentiating such contributions vanish
as $i\lambda \to \infty$), indeed \be Z_{{\cal N}{\cal N}}(L, -L, \lambda) =
i\lambda L + \sqrt{\kappa} \sum_{i=1}^{{\cal N}-1}\int _{-L}^{L} dx\
\psi_{i}(x) W_{i{\cal N}}(x). \label{Z}\ee Due to (\ref{Z}) it is obvious
that in this case it is sufficient to derive the coefficients
$W_{i{\cal N}}$. In any case one can show that the coefficients
$W_{ij}$ satisfy the following equations: \be  && \sum_{ i\neq j}
W_{ij}'E_{ij} -i\lambda \sum_{i\neq {\cal N}} \Big (W_{{\cal N}i} E_{{\cal N}i}
-W_{i{\cal N}}E_{i{\cal N}}\Big ) +\sqrt{\kappa}\sum_{i\neq {\cal N}}\Big (\bar
\psi_{i} W_{{\cal N}i}^2E_{{\cal N}i} + \psi_i W_{i{\cal N}}^2 E_{i{\cal N}} \Big ) =\non\\ &&
\sqrt{\kappa}\sum_{i\neq {\cal N}}\Big (\bar \psi_{i} E_{i{\cal N}} + \psi_{i}
E_{{\cal N}i}\Big ) + \sqrt{\kappa} \sum_{i\neq j,\ i\neq {\cal N},\ j\neq
{\cal N}}\Big (\bar \psi_i W_{{\cal N}j} E_{ij} +\psi_i W_{ij}E_{{\cal N}j} \Big )
\non\\ && - \sqrt{\kappa} \sum_{i\neq j,\ i\neq {\cal N},\ j \neq {\cal N}}\Big
(\bar \psi_{j}W_{{\cal N}j}W_{ij} E_{ij} +\psi_i W_{i{\cal N}} W_{j{\cal N}} E_{j{\cal N}}\Big
). \label{rec1} \ee Finally setting $W_{ij} =\sum_{n=1}^{\infty}
{W_{ij}^{(n)} \over \lambda^n}$ and using (\ref{rec1}) we find
expressions for $W_{i{\cal N}}^{(n)}$ i.e. \be && W_{i{\cal N}}^{(1)}(x) = -i
\sqrt{\kappa} \bar \psi_{i}(x), ~~~~W_{i{\cal N}}^{(2)}(x)=\sqrt{\kappa}
\bar \psi_i'(x) \non\\ && W_{i{\cal N}}^{(3)}(x) = i\sqrt{\kappa}\bar
\psi_i''(x) -i \kappa^{{3\over 2}} \sum_{k} |\psi_{k}(x)|^2\bar \psi_i(x),
~~\ldots. \label{ref2} \ee
In the boundary case we shall need in
addition to (\ref{ref2}) the following objects: \be &&
W_{{\cal N}i}^{(1)} = i\sqrt{\kappa}\psi_i, ~~~~W_{{\cal N}i}^{(2)} = -i
W^{'(1)}_{{\cal N}i} +\sum_{i\neq j,\ i\neq N,\ j\neq N}W^{(1)}_{{\cal N}
j}W_{ji}^{(1)},~~~~W_{ji}^{'(1)} = iW_{j{\cal N}}^{(1)} W_{{\cal N}i}^{(1)}
\non\\ && W_{{\cal N}i}^{(3)} = - iW^{'(2)}_{{\cal N}i}
+W_{i{\cal N}}^{(1)}W_{{\cal N}i}^{(1)}W_{{\cal N}i}^{(1)} + \sum_{i\neq j,\ i \neq {\cal N},
j\neq {\cal N}} W_{{\cal N}j}^{(1)}W_{ji}^{(2)} \non\\ && W_{ij}^{'(2)} =
iW_{i{\cal N}}^{(1)} W_{{\cal N}j}^{(2)}-i W_{j}{\cal N}^{(1)} W_{{\cal N}j}^{(1)}
W_{ij}^{(1)}. \label{reff}\ee
Based on the latter formulas, the
expression (\ref{exp0}), and defining $\hat W(x,\lambda) = W(x, -\lambda)$ we
may rewrite (\ref{sp}) as: \be &&
{\mathbb V}(x,\lambda ,\mu)= {1 \over \lambda -\mu} (1 +W(x))
E_{{\cal N}{\cal N}} (1+W(x))^{-1} + {1 \over \lambda + \mu} (1 + \hat W(x))
E_{{\cal N}{\cal N}} (1+ \hat W(x))^{-1} \non\\ && {\mathbb V}(0, \lambda, \mu)=
(X_0^+)^{-1}\Big ( {1 \over \lambda -\mu} X^+ K^+(\lambda) +   {1
\over \lambda -\mu} K^+(\lambda) X^+ \Big ) \non\\ &&  {\mathbb
V}(0, \lambda, \mu)= (X_0^-)^{-1}\Big ( {1 \over \lambda -\mu}
K^-(\lambda) X^-  +  {1 \over \lambda -\mu} X^- K^-(\lambda)  \Big
)\ee where we define: \be  && X^+ = (1+W(0)) E_{{\cal N}{\cal N}} (1+\hat
W(0))^{-1}, ~~~~X^- =  (1+\hat W(-L)) E_{{\cal N}{\cal N}} (1+W(-L))^{-1} \non\\
&& X_0^+ = \Big [(1+\hat
W(0,\lambda))^{-1}K^+(\lambda)(1+W(0,\lambda))\Big ]_{{\cal N}{\cal N}} \non\\
&& X_0^{-}= \Big [(1+ W(-L,\lambda))^{-1}K^-(\lambda)(1+\hat
W(-L,\lambda))\Big ]_{{\cal N}{\cal N}}. \ee

\end{document}